\documentclass[journal, twocolumn]{IEEEtran}

\usepackage{cite}
\usepackage{graphicx}
\usepackage{units}
\usepackage{textcomp}
\usepackage{color}
\usepackage{listings}

\begin{document}

\title{Control Infrastructure for a Pulsed Ion Accelerator}

\author{A.~Persaud, M.J.~Regis, M.W.~Stettler, V.K.~Vytla %
  \thanks{A.~Persaud, M.J.~Regis, M.W.~Stettler, V.K.~Vytla are with the
  E.O. Lawrence Berkeley National Laboratory, Berkeley, CA 94720, USA (apersaud@lbl.gov)}}

\maketitle

\begin{abstract}
  We report on updates to the accelerator controls for the Neutralized
  Drift Compression Experiment II, a pulsed induction-type accelerator
  for heavy ions. The control infrastructure is built around a LabVIEW
  interface combined with an Apache Cassandra backend for data
  archiving. Recent upgrades added the storing and retrieving of
  device settings into the database, as well as ZeroMQ as a message
  broker that replaces LabVIEW's shared variables. Converting to ZeroMQ
  also allows easy access via other programming languages, such as
  Python.
\end{abstract}


\IEEEpeerreviewmaketitle


\section{Introduction}
Depending on the scale of a project, the control infrastructure and
data archiving systems can range from small desktop applications and
text files to store the data, to complex database driven backends and large frameworks
such as the Experimental Physics and Industrial Control System
(EPICS)\cite{epics}. For the medium-sized ion accelerator described in
this work, we decided to implement a custom solution instead of using
a pre-existing framework. This decision was mainly driven by the
requirement to have a flexible system that can be easily modified (for
example by a scientist with a moderate knowledge of
LabVIEW\cite{labview}). The data archiving system should be able to
handle data acquired over many years. Furthermore, system
configurations should be archived and easy to recall.
We also wanted to take advantage of recent technology developments, such as the
availability of new database systems that can handle big data easily,
scale well, and automate tasks such as synchronization between nodes.

In the following paragraphs, we will give a short overview of the
accelerator and its operation principle. We will then focus
on the control infrastructure, explaining the design goals that resulted
in the hardware and software choices that are used in this experiment,
especially the database backend and the message broker.

The Neutralized Drift Compression Experiment II (NDCX-II) is a high
current, short pulse induction-type linear ion accelerator at the
Lawrence Berkeley National Laboratory (LBNL)\cite{nima-733-226,
  Seidl2016}. After a recent upgrade\cite{Ji2015}, it provides around
$10^{10}$--$10^{11}$ ions per pulse at a beam energy of
\unit[1.2]{MeV} over a pulse duration of \unit[1--2]{ns} in a
millimeter beam spot\cite{Seidl2015}. This corresponds to a peak
current of \unit[$\sim$1]{A}. Ongoing improvements aim to increase the
peak current to around \unit[80]{A}, where intense, short ion pulses
will uniformly heat foils a few microns thick to temperatures of about
\unit[1]{eV}, enabling precision studies of phase transitions in
materials and access to transient states of warm dense matter.
The main concept of the accelerator is to start with a
\unit[$\sim$700]{ns}-long pulsed beam where space charge effects are small,
and then compress the pulses by applying voltage ramps in acceleration
gaps.  These acceleration stages increase the beam energy and also
give the beam a velocity tilt (the design is such that ions arriving
later will receive a higher velocity). Using a drift section after
each acceleration gap results in drift-compression of the beam to
shorter bunch durations. In the \unit[$\sim$10]{m}-long accelerator
there are seven acceleration gaps and seven drift sections that
together compress the beam pulses to a length of \unit[70]{ns} and
give the beam an energy of \unit[$\sim$350]{keV}.
\begin{figure}[ht]
  \centering
  \includegraphics[width=0.9\linewidth]{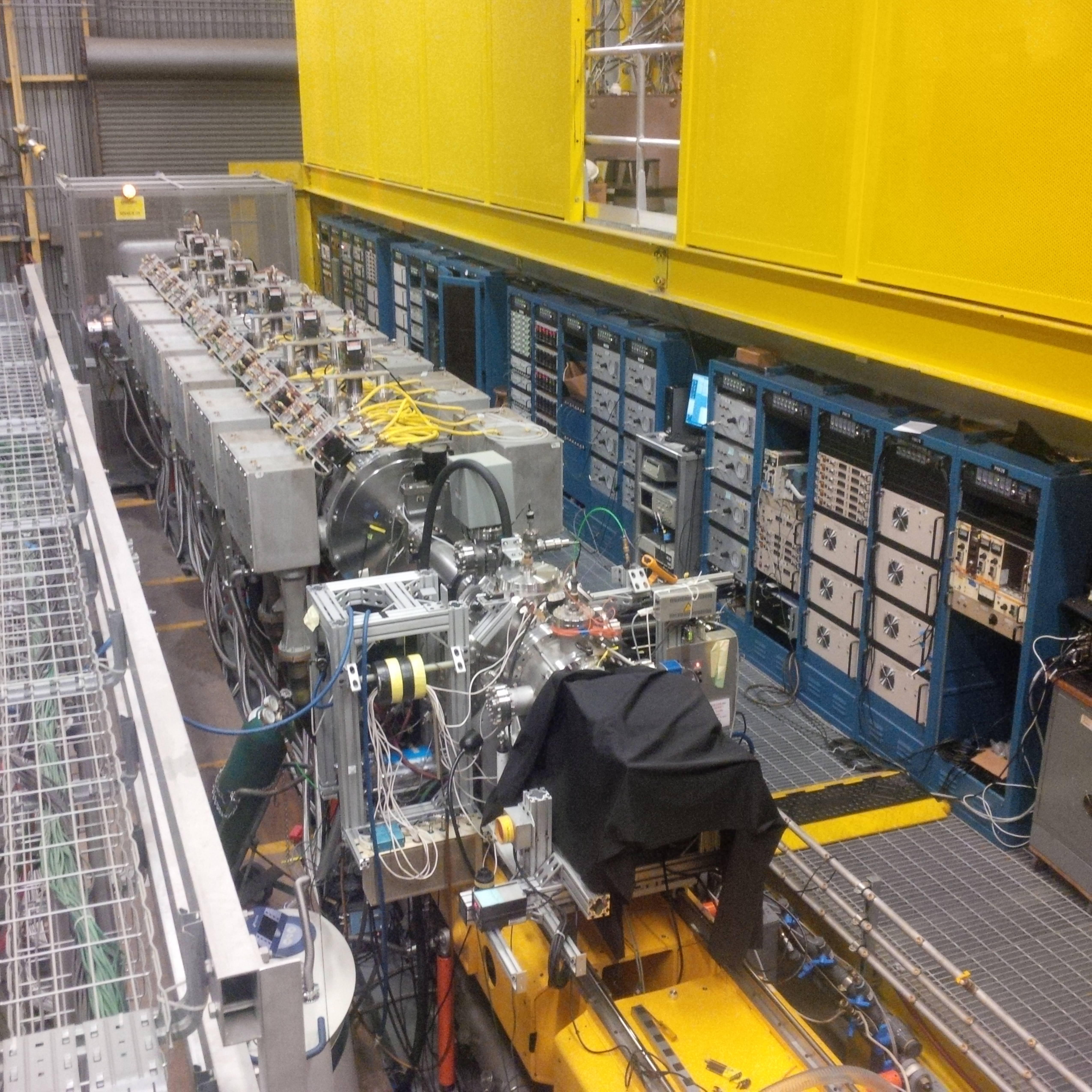}
  \caption{The Neutralized Drift Compression Experiment II. At the
    rear the ion source cage can be seen, followed by the 28
    solenoids and the drift section. At the front, the
    target and diagnostic chamber are visible.}
  \label{fig:beamline}
\end{figure}

Next, there are five Blumlein compressors that provide further
velocity tilts and accelerate the beam to \unit[1.2]{MeV}. At this
point, without an additional element, space charge effects in the beam
would cause its diameter and bunch length to expand rapidly,
prohibiting strong focusing (in time as well as in space). In order to
neutralize the space charge effects, the last Blumlein gap is thus
followed by a plasma-filled drift section. The electrons in the plasma
neutralize the space charge of the beam and allow the velocity tilt to
compress the beam to a bunch length of \unit[1]{ns}.  A strong
focusing (\unit[8]{T}) magnet combined with the neutralized space
charge allows focusing of the beam to a spot size of
\unit[$\sim$1]{mm}. The system also uses 28 solenoid magnets
($\unit[<4]{T}$) along the beamline for beam transport, as well as
inductive beam pickups and beam position monitors. Further diagnostics
are available in the target chamber (Faraday cups, scintillator,
fiber-coupled streak spectrometer, etc.).

The system uses different types of monitors (vacuum gauges,
thermocouples, flow meters, etc.) to continuously measure the status of the accelerator. Furthermore oscilloscopes, precision
trigger systems, and analog
inputs and outputs are used to collect data from pulsed systems involving
the injector, solenoid magnets, and associated diagnostics. A pulse
rate of typically around two shots per minute is implemented,
producing several megabytes of data per shot. The input and output
signals involve approximately 160 oscilloscope channels, 80 delay
pulse generator channels, 50 power supplies, and around 450 analog and
digital I/O ports, see Table~\ref{tab:hardware} for a complete list.
\begin{table*}[ht]
  \renewcommand{\arraystretch}{1.3}
  \centering
  \begin{tabular}{lllcc}
    Manufacturer          &   Model    & Type                   &  Count  & Pulsed/continuous operation \\ \hline \hline
    National Instrument   & PXI-1045   & PXI crate              &    5    & N/A \\
    National Instrument   & PXI-6652   & timing/sync            &    5    & N/A \\
    Greenfield Technology & GFT-9404   & trigger (4+4 channels) &   21    & pulsed \\
    National Instrument   & PXI-5105   & 8-channel oscilloscope &   13    & pulsed \\
    National Instrument   & PXI-5153   & 2-channel oscilloscope &   43    & pulsed \\
    Picotech              & 5443A      & 4-channel oscilloscope &    1    & pulsed \\
    Automation Direct     & Direct Logic 205 & controller       &    5    & continuous \\
    Automation Direct     & DL-260     & CPU                    &    1    & continuous \\
    Automation Direct     & DL-12TR    & relay output module    &    4    & continuous \\
    Automation Direct     & D2-EM      & expansion unit         &    1    & continuous \\
    Automation Direct     & D2-CM      & expansion unit         &    1    & continuous \\
    Automation Direct     & F2-08AD-2  & 8-channel analog input &   21    & continuous \\
    Automation Direct     & H2-EC0M100 & ethernet module        &    7    & continuous \\
    Automation Direct     & D2-32ND3   & 32-channel 24V module  &   34    & continuous \\
    National Instrument   & cDAQ-9188  & CDAQ chassis           &    1    & continuous \\
    National Instrument   & cDAQ-9485  & 8-channel relay module &    4    & continuous \\
    National Instrument   & cDAQ-9213  & 16-channel thermocouple&    3    & continuous \\
    National Instrument   & cFP-1808   & fieldpoint chassis     &    3    & continuous \\
    National Instrument   & RLY-421    & 8-channel relay module &    8    & continuous \\
    National Instrument   & DI-301     & 16-channel 24V input   &    4    & continuous \\
    National Instrument   & AI-112     & 16-channel analog input&    2    & continuous \\
    National Instrument   & AO-210     & 8-channel analog output&    2    & continuous \\
  \end{tabular}
  \caption{List of hardware used for the controls including
    oscilloscopes for diagnostics.}
  \label{tab:hardware}
\end{table*}

\section{Design concept}
Various design choices were made in order to achieve a small, flexible
control system. Industrial I/O ports with ethernet control (Modbus
over Ethernet) and National Instrument (NI) cDAQ were used when
possible for low-bandwith components. For high-bandwidth components we
also used off-the-shelf devices such as PXI modules and standard
oscilloscopes. The use of standard components kept the cost low and
enables easy maintenance and replacements. For control software,
small, user-editable LabVIEW programs were chosen rather than relying
on existing, larger frameworks such as EPICS. This decision was driven
by the fact that a medium-sized accelerator, such as NDXC-II, can
often not afford dedicated software support, and therefore the use of
more complex frameworks seemed prohibitive. Especially, since changes
in the setup over time (for example target chamber diagnostics) were
anticipated from the beginning. For future end-user support of the
facility, scalability of the data archiving system was a key design
objective. To this end, Apache Cassandra \cite{Cassandra} was chosen,
since it is highly scalable and freely available under the Apache
License 2.0. Apache Cassandra is a non-relational database. These types
of databases are also known as NoSQL databases (SQL stands for
Structured Query Language, and is the standard for relational
databases). NoSQL databases often have a simpler design than their
relational counterparts, making scaling and clustering of machines
easier, and often allow for fast read and write operations, as well as
providing the ability to handle very large tables, that is, they can
easily handle data from many shots acquired over many years. NoSQL
databases have seen a large increase in use over the last decade,
accompanied by several new databases such as Cassandra. Since we
only use the database to store data and do not require operations on
the data within the database (e.g. joining tables), we can take
advantage of the benefits a NoSQL database offers.  Cassandra has also
been used by other groups in a similar context\cite{Shen, Kago, Ito}.
Apart from archiving data, we also want to save the configuration
state during each experiment, so that we can easily reload old
configurations of the accelerator. To this end, we save the settings
of power supplies, triggering systems, etc. as well as the settings
for each oscilloscope and other diagnostics for each shot. Since the
configuration data is in principle similar to measured data, we use
the same database to store this information.  ZeroMQ (ZMQ) \cite{ZMQ}
was implemented as a message broker between different LabVIEW and
Python \cite{python} programs. At the beginning, LabVIEW shared
variables were used for this, but switching to ZMQ allows us to also
access our devices using other programming languages, providing an
easy way to script experiments (e.g. parameter scans). A recent report
on utilizing ZMQ for accelerator control is available elsewhere
\cite{Yamashita} and a comparison of ZMQ with other similar software
for accelerator controls has been published by Dworak \textit{et al}. \cite{Dworak}.

\section{Hardware setup}
Nanosecond control of the pulse systems of the accelerator is enabled
using PXI timing modules to trigger the pulsers and the
oscilloscopes. For power supply control, vacuum control, and control
of other critical non-pulsed systems, programmable logic
controllers (PLC) and Industrial I/O have been chosen, thus enabling
the non-pulsed systems to run on their own without a LabVIEW program
or any other desktop computer running. Other non-critical diagnostics
and instruments, such as cameras and stages, are directly controlled
using LabVIEW running on a desktop computer.

\begin{figure}[ht]
  \centering
  \includegraphics[width=0.9\linewidth]{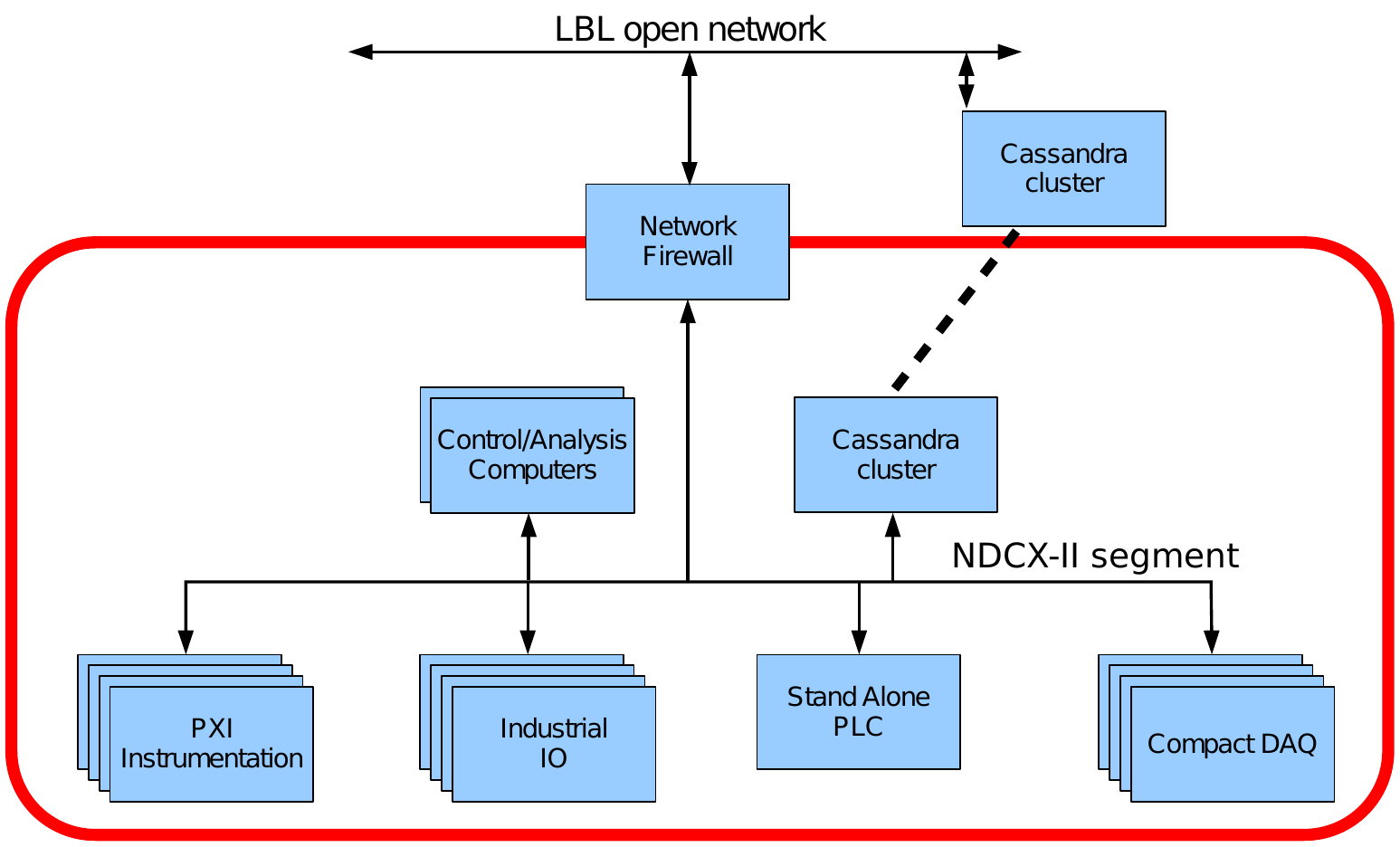}
  \caption{System architecture showing the ethernet connections
    between different components.}
  \label{fig:architecture}
\end{figure}

The entire system is protected by a firewall. A switch
distributes ethernet to each rack via fiber links (to reduce
noise and enable floating of certain racks at high voltage). See
Fig.~\ref{fig:architecture} for a schematic of the architecture layout.

An interlock system used for hardware protection is tied into
the PLC system.

\section{Software setup}
A Cassandra cluster running on Linux virtual machines
is used as the backend to archive data and to read and write settings
for each instrument. Currently the cluster consists of only two
nodes. One node behind the firewall is used as the main instance of
the database. A second node outside the firewall exists to give
read-only access to collaborators, principally to ensure that the main
instance will not have to process any external user requests. It also
creates an automated backup of the data in a different datacenter. If
more resources are needed, additional nodes can be added to Cassandra
within minutes and without shutting down the database. Synchronization
of the data between different nodes is handled automatically by
Cassandra, providing a very low maintenance setup for the database.

Most of the hardware control has been implemented in LabVIEW, due to
pre-existing drivers for the hardware used in the project. Only parts
of the Cassandra binary protocol\cite{cassandraBP3} had to be
implemented to enable reading and writing of data directly to the
database. Our implementation of the binary protocol can be viewed and
downloaded at\cite{cassandra-driver-LV}.

There are two main control computers running LabVIEW programs: one
controlling the pulsed and one controlling the non-pulsed systems. The
computer handling the pulsed system also distributes a global
timestamp for the data acquisition system.

Each device or group of devices (e.g. a PXI crate) is controlled by
LabVIEW code running on a controller or a desktop computer. Each device then
runs its own LabVIEW program that for a pulsed system waits for a hardware
trigger from the timing system. When it receives a trigger the system
acquires the data, then requests a timestamp from the main computer
via ZMQ and saves the data in Cassandra together with its current
settings. It also listens on a different ZMQ port for commands, for
example to fetch and load a different setting from the database. For
non-pulsed systems, data is either acquired and saved to the database
on a fixed time interval (e.g. one second) or the program responsible
for the data acquisition listens on a ZMQ port until a shot has been taken
and then saves its data to the database.

The non-pulsed LabVIEW controls include additional software
interlocks for the user interface, adding some complexity to the
system. The pulsed system on the other hand consists of only a handful
of LabVIEW functions and can easily be adapted to new hardware and diagnostics.

Settings and shot data are being read back from Cassandra, but can also
be sent via ZMQ (bypassing the database).

All data exchanged between LabVIEW and Cassandra is converted from a
LabVIEW-typedef-cluster to a JavaScript Object Notation (JSON)
string\cite{json} for reading and writing using the built-in LabVIEW
functions to handle JSON. JSON allows easy serialization of complex data
structures and also makes it easy to read and write data into the
database using different languages, e.g. Python\cite{python}. JSON has
the advantage that it is a very widespread serialization format
(standard in many web applications), well supported by LabVIEW and
other languages and easy to use and understand. Since the database
uses a compressed storage, we can easily use normal strings for the data
instead of more compact binary notations.

Certain equipment (3 NI-FieldPoint modules) is controlled using the
Generalized Equipment and Experiment Control System (GEECS), a LabVIEW
framework developed at LBNL. Integrating Cassandra data storage and
ZMQ into an existing framework was straightforward.

For data analysis, Python (version 3.5.1) is used to access Cassandra
using the standard cassandra-driver\cite{cassandra-driver-python}. The
LabVIEW clusters map to a dictionary data type in Python
in a straightforward manner using the Python JSON interface.

Git\cite{git} is used as a version control system, using an extension
for better integration with LabVIEW\cite{labviewgit}.

For the interested reader, we made the git repositories containing the
LabVIEW and Python code available online \cite{NDCXII-python, NDCXII-LabVIEW}.

\section{Database layout and setup}
The database layout is kept relatively simple by using only four
different tables, see Tab.~\ref{tab:database-setup}.

\begin{table*}[ht]
  \renewcommand{\arraystretch}{1.3}
  \centering
  \begin{tabular}{lllll}
    Table name & Column name & Column type& Index & Description \\ \hline \hline
    Shot & lvtimestamp & integer & primary key&timestamp when data was taken\\
         & devicename  & text    & clustering key &\\
         & data        & text    & &serialized data\\
         & settinghash & text    & &link to hash in the Settings table\\
         & version     & integer & &version of serialization schema of
                                   the data column\\
    Setting & hash & text & primary key &md5 hash of data column\\
            & data & text & &serialized data\\
            & version & integer & & version of serialization schema of
                                  the data column \\
    Data & devicename & text & primary key& \\
         & date       & integer & primary key& year and month\\
         & eventtime  & timestamp &clustering key& \\
         & value      & float & &\\
    Comments & month   & integer &primary key& year and month\\
             & daytime & integer &clustering key& day and time\\
             & species & text    &&\\
             & comment & text    &&\\
             & sample  & text    &&\\
             & shots   & list of integer && link to lvtimestamp in the
                                           Shot table\\
  \end{tabular}
  \caption{The Apache Cassandra database layout. The data entry for each table
    generally consists of JSON-packed data (versioned) to make the setup more flexible.}
  \label{tab:database-setup}
\end{table*}

For each shot, all data from pulsed devices is saved in the
\textit{Shot} table.  We use a LabVIEW timestamp (in seconds) as the
row key and a device name (e.g. PXI crate number and slot number) to
identify a composite key. The data is then saved as a JSON\cite{json}
string. We also save a version number in case the LabVIEW cluster used
to store the data changes. The settings of each device are saved in a
similar manner in a separate \textit{Setting} table. Here, we use
an md5-hash of the data as the primary key and do not store any device
name or time information. The md5-hash value of the setting string is used
as a key in the Shot table.  Continuous data (e.g. every second or
less) is saved in the \textit{Data} table. For the non-pulsed
data, we only save floating values and a time stamp, and use the
device name as the row key together with a composite key that encodes
the year(Y) and month(M) as an integer in the format YYYYMM. In this
way about 10000 entries are saved in each row, which is a manageable
value for Cassandra. For some diagnostics (e.g. monitoring
temperature) we also set the time-to-live (TTL) so that we have a high
time resolution for recent events, but only a low time resolution for
times when, for example, no shots were taken.

To be able to access groups of shots, we use the fourth \textit{Comments} table, which
saves sets of shot numbers together with some other meta-information
(e.g. sample used) in the database. These are grouped by month, M,  and
year, Y,  in a YYYYMM format and use a composite key of DDHHMMSS
(D=day, H=hour, M=minute, S=seconds) to group a larger amount of items
per row. This way all shots from one experiment or of a certain day
can be grouped together. We also use this table to gather all shots of
a day into a group using a DDHHMMSS=DD420000 key. The creation of this
extra table makes it easy to group shots together or access all shots
from a certain time range, something that can be more difficult in
NoSQL databases.

In our case all of the information in the database is written only
once and afterwards only read back. Table joins, as used in relational
databases, are not necessary. At most only one cross-table lookup is
required, either looking up shots from an entry in the comment
table or looking up settings from a shot. Therefore, a NoSQL database
works very well for our purposes, while providing scalability and
excellent performance.

\section{ZMQ communication}
Communication between devices and the main programs is handled by ZMQ
\cite{ZMQ}. ZMQ provides an easy way to implement network
communication on a basic level. It comes with support for several
often-used network topologies, such as request-reply pairs,
subscriber-publisher, load-balancer, etc. In our case, every device
has the listener of a ZMQ request-reply pair built in, so that it can
receive commands such as \textit{load setting hash xyz} or
\textit{send current data} and reply appropriately. Each device can
also request the timestamp of the current shot using another ZMQ
request-reply pair from the main program that controls the pulsed
system. The timekeeper system will freeze the timestamp for several
seconds when a first request has been made in order to obtain
synchronized timestamps in the database. Since we always have several
seconds between shots, this is an easy solution to supply synchronized
timestamps. The main program also
has a ZMQ publisher running that notifies any ZMQ subscriber when a
shot has been taken. This is used to enable saving of settings and
data from devices that do not receive a hardware trigger, e.g. power
supply controls for charging systems of the pulsed power (we want to
be able to access these settings as well as any measured values
later). Furthermore, we plan on using this feature to automatically
start data analysis or simulation runs for each and every shot.

The use of ZMQ together with a database backend with support for many
programming languages makes it easy to add components that do not
require LabVIEW for data acquisition. For example, we were able to add
a four-channel oscilloscope to a floating high voltage rack using a
small screenless oscilloscope\cite{picotech} that interfaces via USB to a
raspberry-pi computer running a small Python program to take data and
save it to the database. The program implements the same ZMQ channels
and commands as our other LabVIEW data acquisition and therefore
interfaces nicely with our main LabVIEW programs.

\section{Summary and Outlook}
We implemented a control system that uses a NoSQL database (Apache
Cassandra) as a backend and ZMQ as a message broker. Complex data
is saved in JSON format within the database. The combination of JSON
and ZMQ allows us to combine LabVIEW and Python applications across
different computer platforms, operating systems and programming
languages. It also makes it easy to add new capabilities, since any
data can be packaged in a JSON string, and to change the storage
layout without editing the database setting. Operation of the system
described is very reliable.

The layout also allows the system to be scripted using Python (or
other languages) using the ZMQ channels, giving full access to all
hardware settings available through LabVIEW. This enables
simple parameter scans as well as complex multi-parameter scans and
parameter optimizations.

Cassandra is well suited for installation across large clusters
(linear scaling). Similarly, ZMQ has been designed to handle many
connections with very low latency. We therefore believe that a system
based on these components will scale well, for example, for systems
that acquire data at higher frequency or produce more data per
shot. Our system operates at rates of two shots per minute, hence we
have not observed any limitations with respect to the amount
of data produced or the latencies involved.

\section*{Acknowledgments}
This work was supported by the Office of Science of the U.S. Department of Energy under contract
no. DE-AC02-05CH11231.

\bibliographystyle{IEEEtran}
\bibliography{paper}

\end{document}